\documentclass[preprint,amsmath,amssymb,aps,superscriptaddress,showpacs,longbibliography ]{revtex4-1}
\usepackage{times}
\usepackage{mathptmx}
\usepackage{bm}
\usepackage{bbm}
\usepackage{graphicx}
\usepackage[utf8] {inputenc}
\usepackage{natbib}
\usepackage[colorlinks=true,linkcolor=red,urlcolor=blue,citecolor=blue]{hyperref}
\usepackage{comment}

\newcommand{\ket}[1]{\ensuremath{{\vert #1 \rangle}}}

\usepackage{scalerel}
\usepackage{tikz}
\usetikzlibrary{svg.path}
\definecolor{orcidlogocol}{HTML}{A6CE39}
\tikzset{
  orcidlogo/.pic={
    \fill[orcidlogocol] svg{M256,128c0,70.7-57.3,128-128,128C57.3,256,0,198.7,0,128C0,57.3,57.3,0,128,0C198.7,0,256,57.3,256,128z};
    \fill[white] svg{M86.3,186.2H70.9V79.1h15.4v48.4V186.2z}
                 svg{M108.9,79.1h41.6c39.6,0,57,28.3,57,53.6c0,27.5-21.5,53.6-56.8,53.6h-41.8V79.1z M124.3,172.4h24.5c34.9,0,42.9-26.5,42.9-39.7c0-21.5-13.7-39.7-43.7-39.7h-23.7V172.4z}
                 svg{M88.7,56.8c0,5.5-4.5,10.1-10.1,10.1c-5.6,0-10.1-4.6-10.1-10.1c0-5.6,4.5-10.1,10.1-10.1C84.2,46.7,88.7,51.3,88.7,56.8z};
  }
}
\newcommand\orcid[1]{\href{https://orcid.org/#1}{\mbox{\scalerel*{
\begin{tikzpicture}[yscale=-1,transform shape]
\pic{orcidlogo};
\end{tikzpicture}
}{R}}}}

\begin{document}
\title{Nanotube double quantum dot spin transducer for scalable quantum information processing}
\author{Wanlu Song \footnote[3]{co-first authors of this article\label{cf}} }
\email{2017507007@hust.edu.cn}
\author{Tianyi Du \textsuperscript{\ref {cf}}}
\author{Haibin Liu}
\author{Ralf Betzholz\,\orcid{0000-0003-2570-7267}}
\author{Jianming Cai}
\affiliation{School of Physics, International Joint Laboratory on Quantum Sensing and Quantum Metrology, Huazhong University of Science and Technology, Wuhan 430074, China}
\begin{abstract}
One of the key challenges for the implementation of scalable quantum information processing is the design of scalable architectures that support coherent interaction and entanglement generation between distant quantum systems. We propose a nanotube double quantum dot spin transducer that allows to achieve steady-state entanglement between nitrogen-vacancy center spins in diamond with spatial separations up to micrometers. The distant spin entanglement further enables us to design a scalable architecture for solid-state quantum information processing based on a hybrid platform consisting of nitrogen-vacancy centers and carbon-nanotube double quantum dots.
\end{abstract}
%


\date{\today}

\maketitle

\section{Introduction}

Quantum computing with potential revolutionary applications~\cite{Ladd2010,Buluta2011} has raised increasing interest over the past decades and intensive efforts are devoted to the implementation of quantum information processing. A large variety of physical systems provide promising candidates to construct the basic building blocks for quantum-information processing devices, e.g., photons~\cite{Kok2007}, atoms~\cite{Bloch2008}, trapped ions~\cite{Blatt2008}, superconducting circuits~\cite{Clarke2008}, quantum dots~\cite{Loss1998,Hanson2007}, and spins in solids~\cite{Hanson2008}. Despite their individual advantages, each of these physical systems is accompanied by its own drawbacks. These shortcomings call for the development of hybrid quantum systems~\cite{Wallquist_2009_PST137,Pirkkalainen_2013_N494,Xiang_2013_RMP85,Kurizki_2015_PotNAoS112,LiPB2016} that combine the advantages of its constituents in order to overcome the difficulties toward the implementation of powerful quantum information processing devices. One well recognized
severe challenge in terms of scalability is the implementation of coherent coupling between spatially separated quantum systems.

Nitrogen-vacancy (NV) centers in diamond consist of both an electron spin and an intrinsic nuclear spin, where the electron spins can serve as a register to process quantum information, due to their excellent coherent controllability~\cite{Neumann2010_NP249}, and the nuclear spins can serve as a memory to store quantum information, due to their superb coherence time~\cite{Maurer2012}. Unfortunately, the prospect of using NV centers for scalable quantum computing is hindered by the fact that the direct coupling between NV centers decays rapidly with their distance \cite{Dol_13_NP}. In order to overcome this obstacle, several schemes have been proposed using microwave and optical cavities~\cite{Kub_10_PRL,Kub_11_PRL,Zhu_11_NL,Englund2010,Wolters2010,Sar2011}, mechanical oscillators~\cite{Bennett2013,Rab_10_NP,XuZY2009,ZhouLG2010,Chotor2013,LiPB2016,CaoPH2017,CaoPH2018}, spin-photon interface \cite{Ber_13_NL} to mediate the coupling between distant NV centers. However, the goal of long-range coupling between solid-state spins in a deterministic and scalable manner remains challenging to achieve due to e.g. the influence of cavity losses and the thermalization of mechanical oscillators.
In this work, we propose an efficient strategy for steady-state entanglement generation between NV-center electron spins at micrometer distances, which is mediated by the leakage current of a carbon-nanotube double quantum dot (DQD)~\cite{Laird2015,Rohling2012,Rasmussen2012,PeiF2012}. Each quantum dot locally interacts with a single NV center in the proposed hybrid platform. Due to the Pauli exclusion principle, we find that the NV-center electron spins will be driven into a maximally entangled state along with the electrons being blocked in the DQD. The scheme requires only voltage control of the nanotubes and microwave driving of the NV-center electron spins, which is feasible within current state-of-the-art experimental capabilities. In addition, the steady-state entanglement of the electron spins can be exploited to realize an entangling gate between the nuclear spins associated with the NV centers via the hyperfine coupling. Therefore, the hybrid platform allows to generate nuclear-spin cluster states~\cite{Nemoto2014_PRX} for universal measurement-based quantum computation~\cite{RN3} with excellent scalability, and provides a new route toward solid-state quantum information processing.
%

\begin{figure}[t]
  \hspace{-0.3cm}
  \includegraphics[width=7.5cm]{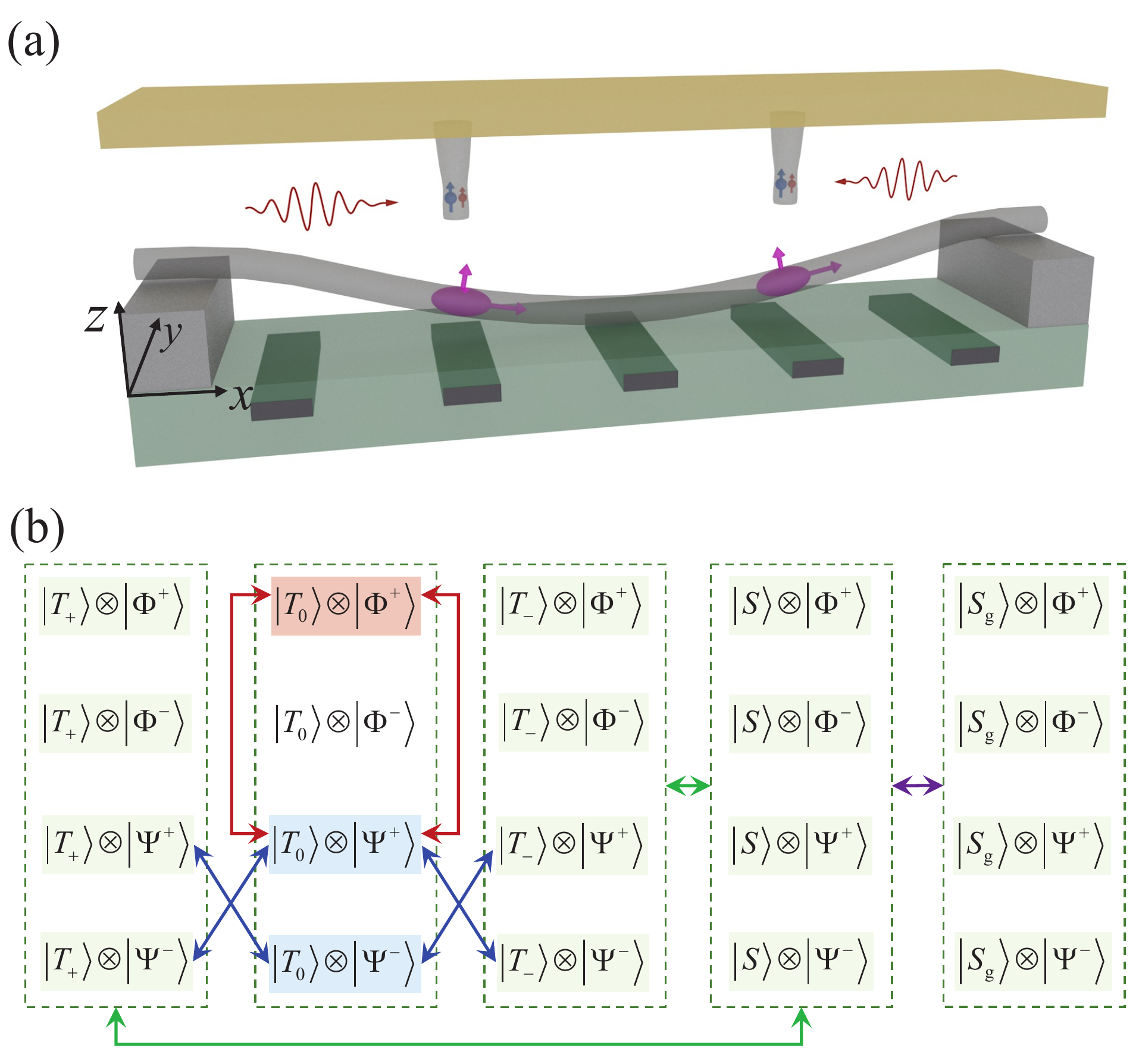}
  \caption{{\bf (a)} Schematic of the hybrid building block: diamond pillars (transparent gray) containing single NV centers placed above  carbon nanotubes bridging on the source and drain contacts. The gates that create the DQD (magenta) are below the nanotube \cite{PeiF2012,Laird2013}. The DQD confined in the nanotube is used to mediate the generation of steady-state entanglement between the electron spins (blue) of two driven NV centers, where diamond nanopillars can be located close to the DQD using the technology developed for diamond scanning probes, see e.g. Ref.~\cite{Maletinsky_2012_NN7,Schaefer-Nolte_2014_RoSI85,Gro_17_NL}. {\bf (b)} Transition diagram of the DQD and NV-center electron
spins hybrid system in the Hilbert space spanned by $\left\{\left\vert T_{+}\right\rangle,\left\vert
T_{0}\right\rangle,\left\vert T_{-}\right\rangle,\left\vert S\right\rangle,\left\vert S_{\text{g}}\right\rangle\right\}$ (DQD) and $\left\{\left\vert\Phi^{+}\right\rangle,\left\vert\Phi^{-}\right\rangle,\left\vert\Psi^{+}\right\rangle,\left\vert \Psi^{-}\right\rangle\right\}$ (NV centers). Arrows represent the main transition channels induced by the external
  magnetic field (green), the microwave driving field (red), the magnetic dipolar coupling (blue), and the tunnelling of the DQD (purple). The state $\left\vert T_{0}\right\rangle\otimes\left\vert\Phi^{-}\right\rangle$ represents the only decoupled state from tunnelling.}\label{model}
\end{figure}

\section{Model}

The hybrid system we propose consists of diamond pillars and carbon nanotubes. Single NV centers are embedded in the diamond pillars, containing both an electron and a nuclear spin. Each carbon nano\-tube bridges on the source and drain contacts, and electrons can be confined by the gate voltage to form a DQD.  We start by considering a single building block as shown in Fig.\ref{model}(a), i.e., a hybrid system consisting of two distant NV-center electron spins, each of which locally couples to a quantum dot.
%

%
In the $j$th NV-center electron spin ($j=\text{L, R}$), one can encode a qubit in the spin sublevels $
\left\vert0\right\rangle_{j}\equiv\left\vert m_{\text{s}}=+1\right\rangle_{j}$ and $\left\vert1\right\rangle_{j}
\equiv\left\vert m_{\text{s}}=-1\right\rangle_{j}$ of the ground-state manifold, which can be coherently driven by optical
adiabatic-passage control~\cite{Golter2014112}. This
induces two dressed qubits robust against decoherence as described by the Hamiltonian~\cite{SI}
\begin{equation}\label{Hnv}
  \hat{H}_{\text{es}}=\sum_{j=\text{L},\text{R}}\frac{\hbar\Omega_{j}}{2}\hat{s}_{x}^{\left(j\right)},
\end{equation}
where $\hat{\mathbf{s}}_{j}=\big(\hat{s}_{x}^{\left(j\right)},\hat{s}_{y}^{\left(j\right)},\hat{s}_{z}^{\left(j\right)}\big)$
are the Pauli vectors and $\Omega_{j}$ are the effective Rabi frequencies. For later use, we define the four Bell states of the two NV-center electron-spin system as $|\Phi^\pm\rangle=\left(|0\rangle_{\text{L}}|
0\rangle_{\text{R}}\pm|1\rangle_{\text{L}}|1\rangle_{\text{R}}\right)/\sqrt{2}$ and $|\Psi^\pm\rangle=\left(|
0\rangle_{\text{L}}|1\rangle_{\text{R}}\pm|1\rangle_{\text{L}}|0\rangle_{\text{R}}\right)/\sqrt{2}$. On the other
hand, in the $j$th quantum dot ($j=\text{L, R}$), one can encode a valley-spin qubit in one of the Kramers
doublets~\cite{Laird2013} formed by the states $\left\vert\Uparrow\right\rangle_{j}$ and $
\left\vert\Downarrow\right\rangle_{j}$~\cite{SI}. Under a magnetic field $\mathbf{B}$, their respective Hamiltonians read~\cite{Szechenyi2015,Szechenyi2015,Szechenyi2017,SI}
\begin{equation}\label{Hvs}
  \hat{H}_{\text{vs}}^{\left(j\right)}=\frac{\mu_B}{2}\mathbf{B}_{\text{eff}}^{(j)}
  \cdot\hat{\mathbf{v}}_{j},
\end{equation}
where $\hat{\mathbf{v}}_{j}=\left(\hat{v}_{x}^{\left(j\right)},\hat{v}_{y}^{\left(j\right)},\hat{v}_{z}^{\left(j\right)}\right)$ are the Pauli vectors and $\mu_{B}$ is the Bohr magneton. The effective magnetic fields acting on the valley-spin qubits are given by $\mathbf{B}_{\text{eff}}^{(j)}=\mathbf{g}_{j}\cdot\mathbf{B}$, with the anisotropic $g$ tensors \begin{equation}\label{g}
  \mathbf{g}_{j}=
  \begin{pmatrix}
    g_{\parallel}\cos^{2}{\alpha_{j}}+g_{\perp}\sin^{2}{\alpha_{j}} & 0 & \left(g_{\parallel}-g_{\perp}\right)\sin{\alpha_{j}}\cos{\alpha_{j}} \\
    0 & g_{\perp} & 0 \\
    \left(g_{\parallel}-g_{\perp}\right)\sin{\alpha_{j}}\cos{\alpha_{j}} & 0 & g_{\parallel}\sin^{2}{\alpha_{j}}+g_{\perp}\cos^{2}{\alpha_{j}}
  \end{pmatrix},
\end{equation}
where $g_{\parallel}$ and $g_{\perp}$ are the local principal values and $\alpha_{j}$ is the angle between the
principal axis of $g_{\parallel}$ and the $x$-axis.

Due to the Coulomb blockade~\cite{Kouwenhoven2001,Hanson2007}, under a large bias voltage, the
electrons in the carbon nanotube transport from the source to the drain through the DQD via the cycle $
\left(0,1\right)\rightsquigarrow\left(1,1\right)\leftrightarrow\left(0,2\right)\rightsquigarrow\left(0,1\right)$, where $
\left(n_{\textbf{L}},n_{\textbf{R}}\right)$ represents the numbers of confined electrons in the left and right
quantum dots. However, when the two electrons in the $\left(1,1\right)$ configuration occupy one of the triplet-like
states $\left\vert T_{0}\right\rangle$ or $\left\vert T_{\pm}\right\rangle$~\cite{Hanson2007}, given by $\left\vert
T_{0} \right\rangle = \left( \left\vert \Uparrow \right\rangle_{\text{L}} \left\vert \Downarrow \right\rangle_{\text{R}}
+ \left\vert \Downarrow \right\rangle_{\text{L}} \left\vert \Uparrow \right\rangle_{\text{R}} \right)/\sqrt{2}$, $
\left\vert T_{+} \right\rangle = \left\vert \Uparrow \right\rangle_{\text{L}} \left\vert \Uparrow \right\rangle_{\text{R}}
$, and $\left\vert T_{-} \right\rangle = \left\vert \Downarrow \right\rangle_{\text{L}} \left\vert \Downarrow
\right\rangle_{\text{R}}$, the $\left(1,1\right)\rightarrow\left(0,2\right)$ transition is forbidden due to the Pauli
exclusion principle. In such a Pauli-blockade regime~\cite{Szechenyi2015}, the spin-conserving tunneling
between the two quantum dots can be described by
\begin{equation}
  \hat{H}_{\text{t}}=\hbar J\left(\left\vert S_{\text{g}}\right\rangle\left\langle S\right\vert +\left\vert S\right\rangle\left\langle S_{\text{g}}\right\vert\right),
\end{equation}
with the tunneling rate $J$ between the singlet-like states $\left\vert S \right\rangle =  \left( \left\vert \Downarrow
\right\rangle_{\text{L}} \left\vert \Uparrow \right\rangle_{\text{R}} - \left\vert \Uparrow \right\rangle_{\text{L}}
\left\vert \Downarrow \right\rangle_{\text{R}} \right)/\sqrt{2}$ in the $\left(1,1\right)$ configuration and the
corresponding singlet-like state $\left\vert S_{\text{g}}\right\rangle$ in the $\left(0,2\right)$ configuration. We remark that the tunneling rate $J$ can reach 100 MHz for an inter-dot distance of a micrometer \cite{Bie_05_xx}. In the case
$\alpha_{\text{R}}=-\alpha_{\text{L}}=\alpha$ and $\mathbf{B}=\left(0,0,B_{z}\right)$, on which we will focus, $|
T_\pm\rangle$ couple with the singlet-like state $|S\rangle$, while  $|T_{0}\rangle$ remains
blocked~\cite{Szechenyi2015,Szechenyi2017}. Assisted by such a $T_0$-blockade mechanism, as well as the
dipole-dipole coupling between the NV centers and the quantum dots, a maximally entangled steady state of
the NV-center electron spins can be achieved.

\section{Steady-state entanglement}

To illustrate the essential idea of our proposal, we first concentrate on the $\left(1,1\right)$ configuration. The total Hamiltonian in this subspace includes the part Eq.~\eqref{Hnv}  for
the NV-center electron spins and Eq.~\eqref{Hvs} for the valley-spin qubits, as well as their magnetic dipole-dipole interaction, which can be written as
\begin{equation}\label{HII1}
    \hat{H}_{\left(1,1\right)}=\hat{H}_{\text{es}}+\sum_{j=\text{L},\text{R}}\left[\epsilon\tilde{\mathbf{n}}_{j}\cdot\hat{\mathbf{v}}_{j} -\kappa_j(\tilde{\mathbf{n}}_j\cdot\hat{\mathbf{v}}_{j})\hat{s}_{z}^{(j)}\right],
\end{equation}
where we introduce the two vectors $\tilde{\mathbf{n}}_\text{L}=\left(-\eta,0,\xi\right)$ and $\tilde{\mathbf{n}}_\text{R}=\left(\eta,0,\xi\right)$,
with $\xi=g_{\parallel}\sin^{2}{\alpha}+g_{\perp}\cos^{2}{\alpha}$ and $\eta=\left(g_{\parallel}-g_{\perp}\right)\sin{\alpha}\cos{\alpha}$. Furthermore, we defined $\epsilon=\mu_BB_z/2$ and the dipole-dipole coupling
strength $\kappa_{j}=\mu_{0}\mu_{B}^{2}g_{\text{s}}/(4\pi r^{3}_{j})$, where $g_{\text{s}}$ is the electron $g$ factor and $r_{j}$ represents the distance between the $j$th NV center and the $j$th quantum dot. We assume
$r_{\text{L}}=r_{\text{R}}$ (which leads to the condition $\kappa_{\text{L}}=\kappa_{\text{R}}=\kappa$) by pulsed Hamiltonian
engineering~\cite{SI}. In this case, with the further condition $\Omega_{\text{L}}=\Omega_{\text{R}}
=\Omega$, it can be seen that under Hamiltonian Eq.~\eqref{HII1} only the state $\left\vert
T_{0}\right\rangle\otimes\left\vert \Phi^-\right\rangle$ is uncoupled from the other basis states, see Fig.~\ref{model}(b).
%

\begin{figure}[t]
  \centering
  \includegraphics[width=8cm]{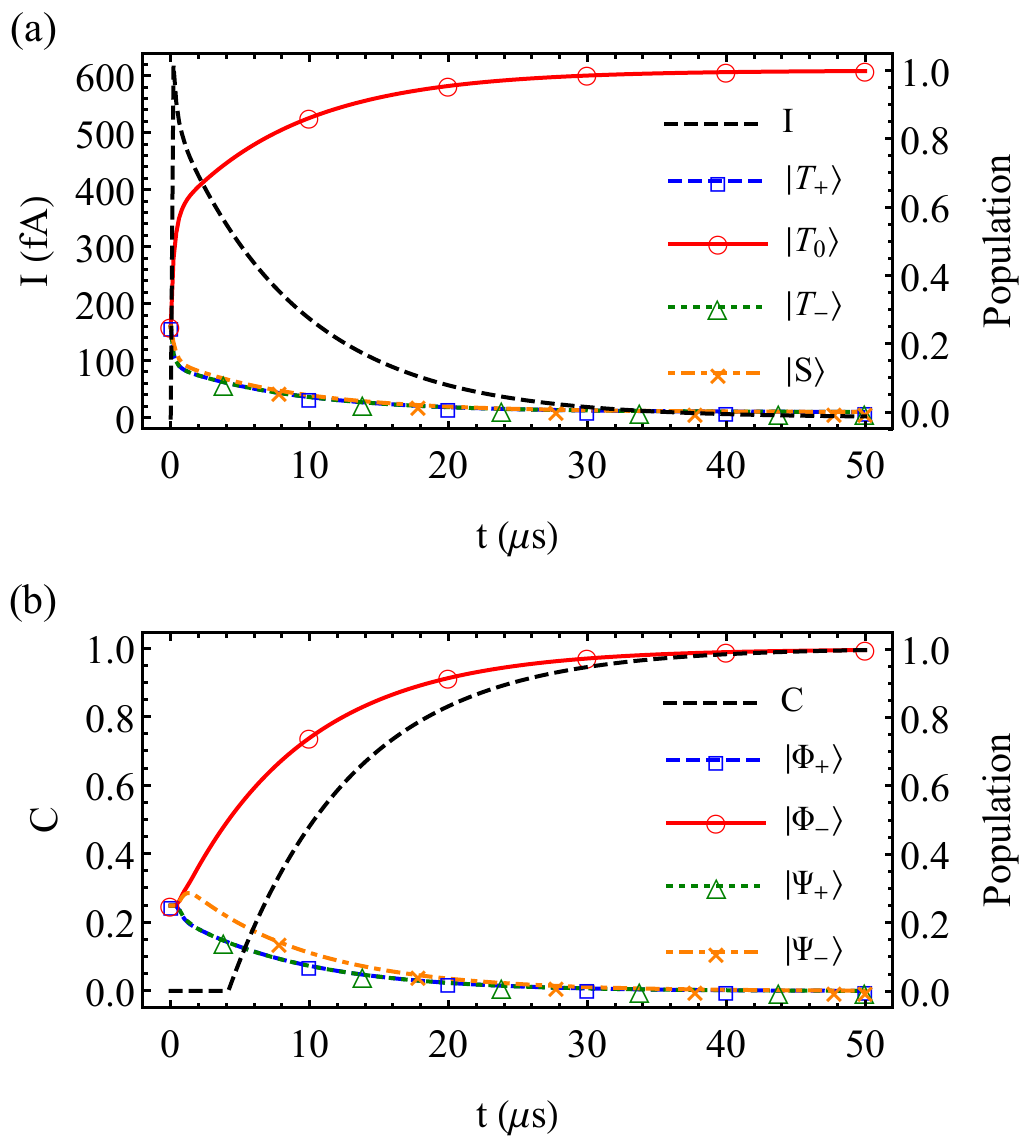}
  \caption{({\bf a}) Leakage current $I$ and the populations of $\left\vert T_{+}\right\rangle$, $\left\vert
  T_0\right\rangle$, $\left\vert T_{-}\right\rangle$, and $\left\vert S\right\rangle$ in the DQD. ({\bf b}) Concurrence $C$ and the populations of Bell states $\left\vert \Phi^\pm\right\rangle$ and $\left\vert
  \Psi^\pm\right\rangle$ in the NV-center electron spins. The parameters are $\Omega/2\pi=0.6$ MHz, $J/
  2\pi=24$ MHz, $\Gamma_{\text{in}}=\Gamma_{\text{out}}=0.5$ GHz, $\Delta=0$, $g_{\parallel}=30$,
  $g_{\perp}=1$, $B_{z}=5$ mT, $\alpha=\pi/36$, and $r_{\text{L}}=r_{\text{R}}=6$ nm.} \label{Fig2}
\end{figure}

%
The underlying mechanism can be understood under the following considerations. The external magnetic field,
couples the states $\left\vert T_{\pm}\right\rangle\otimes\left\vert \phi\right\rangle$ to the states $\left\vert
S\right\rangle\otimes\left\vert \phi\right\rangle$ at rate $\epsilon$, with $\left\vert\phi\right\rangle\in\{\left\vert
\Phi^\pm\right\rangle,\left\vert \Psi^\pm\right\rangle\}$. Thus the states $\left\vert
T_{0}\right\rangle\otimes\left\vert \phi\right\rangle$ are uncoupled with the other basis states in the absence of
the NV centers. However, in their presence, the dipole-dipole coupling, gives rise to transitions between the
states $\left\vert T_{0}\right\rangle\otimes\left\vert\Psi^\mp\right\rangle$ and $\left\vert T_{\pm}\right\rangle\otimes\left\vert \Psi^\pm\right\rangle$ at rate $\kappa$. In addition, the coherent driving of the NV-
center electron spins, couples all states involving $\left\vert \Phi^+\right\rangle$ and $\left\vert \Psi^+\right\rangle$ at rate $\Omega$. This shows that $\left\vert T_{0}\right\rangle\otimes\left\vert \Phi^-
\right\rangle$ is the unique decoupled state in this setting and is eventually reached by the system's dynamical evolution from any initial condition, as we will show in the following more detailed analysis.
The time evolution of the density operator $\hat\rho$, combining the DQD in the $(0,1)$, $(1,1)$, and $(0,2)$
subspaces and the NV-center electron spins, can be described by the quantum-transport master equation $\partial\hat\rho/\partial t=[\hat{H}_{\left(0,1\right)}\oplus\left(\hat{H}_{(1,1)}\oplus\hat{H}_{\left(0,2\right)}+\hat{H}_t\right),\hat\rho]/i\hbar+\mathcal{L}\hat\rho$~\cite{Gurvitz1996,LiXQ2005}, with
\begin{eqnarray}\label{}
  &&\hat{H}_{\left(0,1\right)}=\hat{H}_{\text{es}}+\epsilon\tilde{\mathbf{n}}_{\text{R}} \cdot\hat{\mathbf{v}}_{\text{R}}-\kappa_\text{R}(\tilde{\mathbf{n}}_\text{R}\cdot\hat{\mathbf{v}}_{\text{R}})\hat{s}_{z}^{(\text{R})},\\
  &&\hat{H}_{\left(0,2\right)}=\hat{H}_{\text{es}}+\Delta\left\vert S_{\text{g}}\right\rangle\left\langle S_{\text{g}}\right\vert,\label{eq:Delta}
\end{eqnarray}
and the superoperator
\begin{eqnarray}
  \mathcal{L}\hat\rho&=&\sum_{\psi}\left[\frac{\Gamma_{\text{in}}}{2}\left(2\hat{a}_{1\psi}^{\dag}\hat\rho\hat{a}
  _{1\psi} -\hat{a}_{1\psi}\hat{a}_{1\psi}^{\dag}\hat\rho-\hat\rho\hat{a}_{1\psi}\hat{a}_{1\psi}^{\dag}\right)\right.
  \nonumber\\
&&+\left.\frac{\Gamma_{\text{out}}}{2}\left(2\hat{a}_{2\psi}\hat\rho\hat{a}_{2\psi}^{\dag}-\hat{a}_{2\psi}^{\dag}\hat{a}_{2\psi}\hat\rho-\hat\rho\hat{a}_{2\psi}^{\dag}\hat{a}_{2\psi}\right)\right].
\end{eqnarray}
Here, $\Delta$ is the energy difference between the two singlet-like states $\left\vert S\right\rangle$ and $\left\vert
S_{\text{g}}\right\rangle$. Furthermore, $\hat{a}_{1\psi}^{\dag}$ is the creation operator representing the
injection of an unpolarized electron from the $\left(0,1\right)$ to the $\left(1,1\right)$ configuration with rate $\Gamma_{\text{in}}$ and $\hat{a}_{2\psi}$ is the annihilation operator representing the ejection of an
unpolarized electron from the $\left(0,2\right)$ to the $\left(0,1\right)$ subspace with rate $\Gamma_{\text{out}}$, where $\left\{\left\vert\psi\right\rangle\right\}$ can be any complete and orthogonal set of states of the valley-
spin qubits. In order to investigate the time evolution of the hybrid system under this dynamics in more detail,
we assume that the whole system is initially in a completely mixed state. The dynamical behavior of the DQD
degree of freedom can be characterized by the leakage current $I=e\Gamma_{\text{out}}\sum_{\psi}\text{Tr}
\big[\hat{a}_{2\psi}^{\dag}\hat{a}_{2\psi}\hat\rho\big]$~\cite{Gurvitz1996,LiXQ2005}, with the elementary charge $e$, and the populations of its states. The NV-center electron spins, on the other hand, can be
characterized by the concurrence $C$ (i.e., a renowned measure of two-qubit
entanglement~\cite{Wootters1998}) and the populations of the four Bell states. Figure~\ref{Fig2}(a) and (b) show these quantities for the DQD and the NV-center electron spins, respectively. Due to the external
magnetic field, the fast transition from $\left\vert T_{\pm}\right\rangle\otimes\left\vert\phi\right\rangle$ to $\left\vert S\right\rangle\otimes\left\vert\phi\right\rangle$ at rate $\sqrt{2}\epsilon\eta/\hbar$ leads to a rapid
increase in the leakage current. The speed of entanglement generation is mainly determined by the transition
rate ($\gamma_{\kappa}=\sqrt{2}\kappa\eta/\hbar$) from $\left\vert T_{0}\right\rangle\otimes\left\vert \Psi^\mp\right\rangle$ to $\left\vert T_{\pm}\right\rangle\otimes\left\vert\Psi^\pm\right\rangle$ arising from the
magnetic dipole-dipole coupling between NV-center electron spins and DQD. Our numerical simulations
suggest that the entanglement generation is most efficient by choosing $\Omega\simeq \gamma_{\kappa}$~\cite{SI}. During the tunneling process, the population of the state $\left\vert
T_{0}\right\rangle\otimes\left\vert \Phi^-\right\rangle$ becomes dominant due to the fact that this state is the only state decoupled from the tunneling dynamics, i.e., it is a dark state with respect to the leakage current.
Eventually, the leakage current decreases to zero and the two NV-center electron spins are prepared into the
maximally entangled state $\left\vert \Phi^-\right\rangle$.
\section{Performance under noise}
\begin{figure}[t]
  \centering
  \includegraphics[width=8cm]{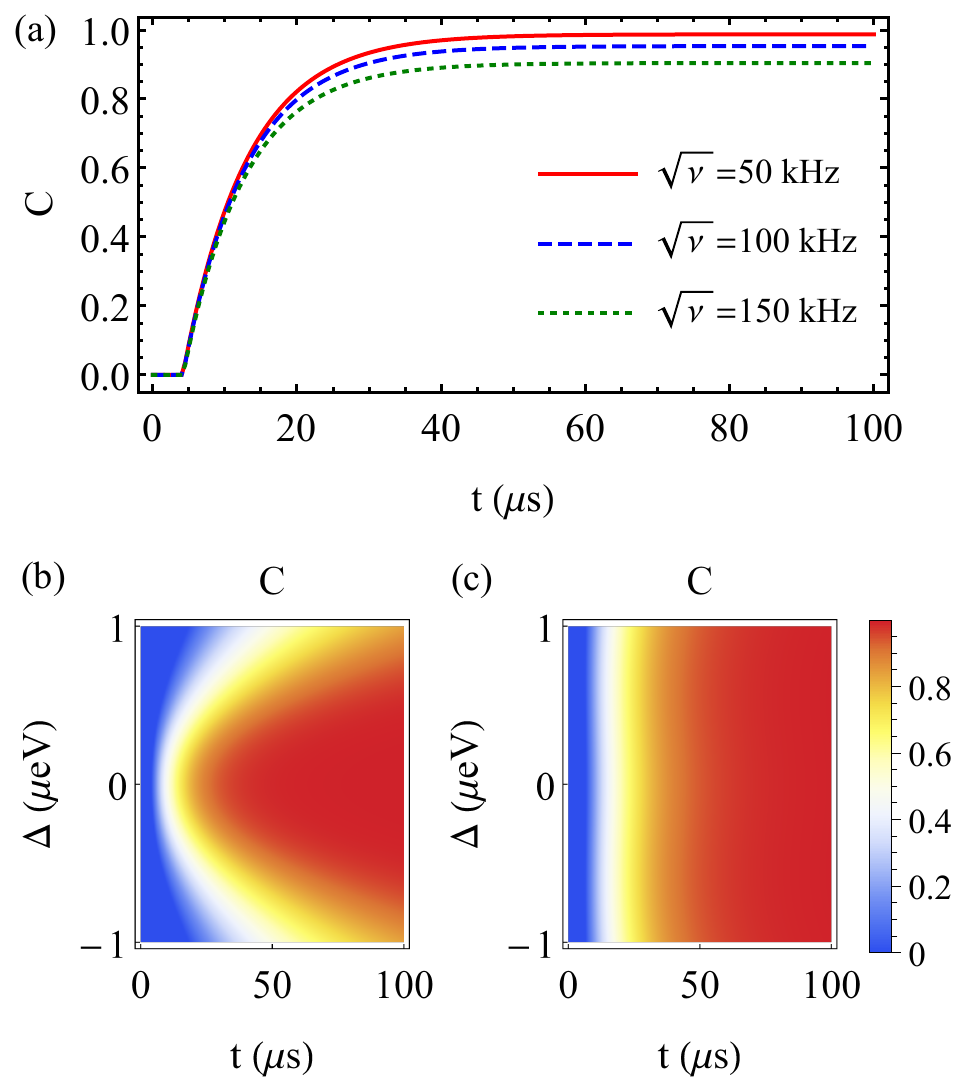}
  \caption{Influence of noise on the generation of steady-state entanglement. ({\bf a}) Concurrence $C$ of two NV-center electron spins under the influence of magnetic field fluctuations with a variance $\nu$. The
 parameters are the same as in Fig.~\ref{Fig2}. ({\bf b-c}) Concurrence $C$ as a function of the energy shift $\Delta$ for different parameters: ({\bf b}) $\Gamma_{\text{in}}=\Gamma_{\text{out}}=0.5$ GHz and $J/2\pi= 24$ MHz; ({\bf c}) $\Gamma_{\text{in}}=\Gamma_{\text{out}}=2$ GHz and $J/2\pi= 36$ MHz. The other parameters are the ones from Fig.~\ref{Fig2}.}\label{Fig3}
\end{figure}

We proceed to investigate the influence of noise in the hybrid system on the generation of steady-state entanglement. The natural abundance of $^{13}\mbox{C}$ (carrying a nuclear spin-1/2) in diamond leads to the fact that the energy levels of the NV-center electron spin are affected by the surrounding $^{13}\mbox{C}$ nuclear spins. The effect of such a nuclear-spin bath can be modeled by a random magnetic field acting on the electron spins, which fulfills a zero-mean Gaussian distribution $\exp(-\delta_{j}^{2}/2\nu)/\sqrt{2\pi\nu}$. Results obtained for different noise variances $\nu$ are shown in Fig.~\ref{Fig3}(a). It can be seen that the magnetic field fluctuations degrade the steady-state entanglement of the NV-center electron spins, however, a highly entangled steady state can still be achieved. The influence of magnetic field fluctuations can be efficiently reduced using dynamical decoupling and isotopically engineered diamond~\cite{Ryan2010,Naydenov2011,Balasubramanian2009}. In addition, by using the NV-center nuclear spins, it is possible to realize entanglement purification~\cite{Hanson2017} in order to prepare a maximally entangled final state.
%

For the carbon nanotubes, an isotopically purified fabrication allows for devices with very few nuclear spins~\cite{Laird2015,Bulaev2008,Rudner2010}. However, as the carbon nanotube DQD is controlled by the
applied gate voltage~\cite{Mason2004}, voltage noise will lead to electric potential fluctuations and thereby
energy level shifts of the DQD, i.e., the parameter $\Delta$ in Eq.~\ref{eq:Delta}. In the presence of this
electric noise, the Coulomb blockade is fragile while the Pauli blockade remains highly
robust~\cite{Szechenyi2017}. In this sense, the effect of electric potential fluctuations is  relatively weak in our
scheme, since it relies only on the Pauli blockade. To demonstrate this influence in detail, we investigate the
role of an energy difference $\Delta$ between the states $\left\vert S\right\rangle$ and $\left\vert S_{g}
\right\rangle$. As shown in Fig.~\ref{Fig3}(b), the non-zero energy difference slows down the entanglement
generation, as compared to the resonant case $\Delta=0$. However, this can be compensated by choosing
proper values of the injection and ejection rates as well as the tunneling rate~\cite{SI}. Here, under optimized
conditions, we find that the dynamical behavior of the entanglement is tolerant against energy differences $
\Delta$ varying from $-1$ $\mu$eV to $1$ $\mu$eV, as shown in Fig.~\ref{Fig3}(c). We point out that the
electric potential fluctuations can be reduced below this level by improved device fabrication~\cite{Jiang2016}.

\begin{figure}[t]
    \includegraphics[width=8cm]{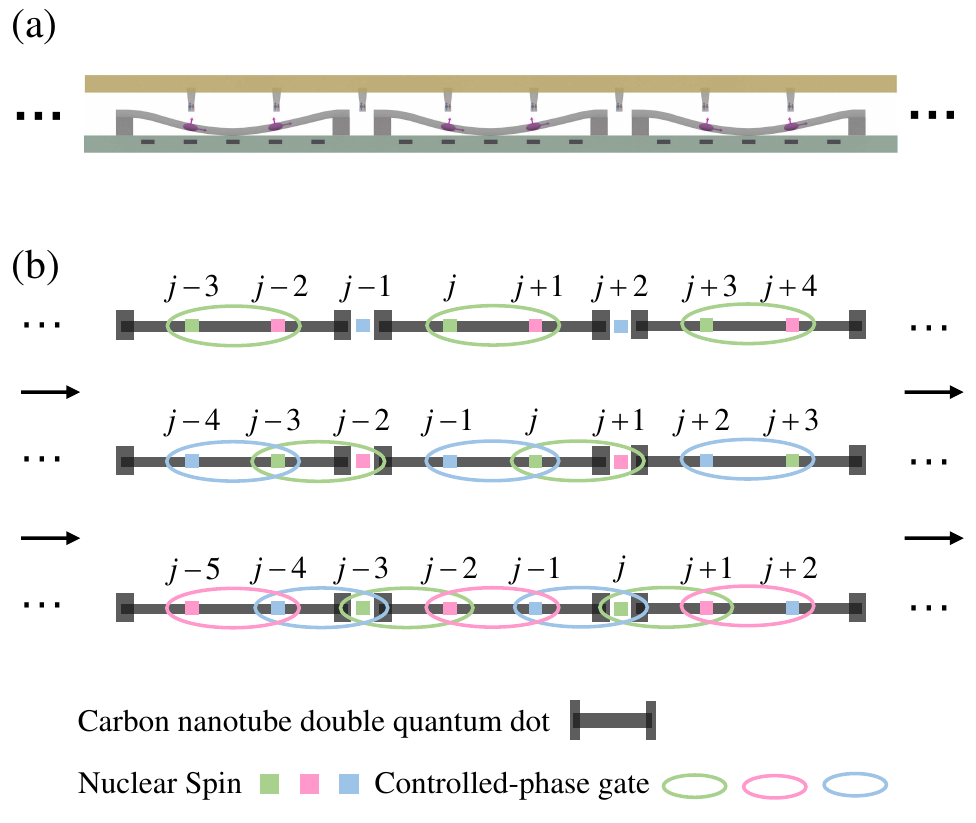}
    \caption{Scalable architecture for multi-qubit entanglement generation. ({\bf a}) One-dimensional (1D) array of hybrid building blocks as shown in Fig.~\ref{model} (a). ({\bf b}) Top row: Controlled-phase gates are implemented between each pair of nuclear spins (in green ellipses) using the DQD-mediated entangled NV
center electron spins; Shifting the pillars and implementing the controlled-phase gate on each alternate pair of nuclear spins (in blue and pink ellipses) subsequently (middle and bottom row) allows to prepare 1D cluster state.
}
    \label{Fig4}
\end{figure}

%
\section{Scalable multi-qubit entanglement}

Once the maximally entangled state $\left\vert\Phi^{-}\right\rangle$ of the NV-center electron-spin pairs is prepared via the coupling to the DQD, we can exploit this
entanglement to realize a controlled-phase gate between $^{15}$N nuclear spins associated with the NV centers. This is achieved based on the hyperfine coupling described by the Hamiltonian~\cite{SI}
\begin{equation}
    \hat{H}_{\text{hf}}=\sum_{j=\text{L,R}}A_{\parallel}\hat{s}_{z}^{\left(j\right)}\hat{I}_{z}^{\left(j\right)},
\end{equation}
with the coupling strength $A_{\parallel}/2\pi=3.03$ MHz~\cite{Felton2009} and the spin-1/2 operator $\hat{I}_{z}^{\left(j\right)}$ of the $j$th nuclear spin. The transversal coupling is safely neglected due to the large
energy mismatch. We remark that a controlled-phase gate, as an entangling gate, together with single-qubit rotations form a set of universal quantum gates.
The controlled-phase gate between$^{15}$N nuclear spins can be realized with the following four steps: (i) A $\pi/4$-$\hat{x}$ rotation on the left electron spin; (ii) Coherent evolution governed by the hyperfine interaction
for a time $t=\pi/2A_{\parallel}$; (iii) A $\pi/4$-$\hat{x}$ rotation on both electron spins; (iv) Measurement of the NV-center electron spins in the $x$-basis ($\{\ket{+}$, $\ket{-}\}$), resulting in an effective controlled-phase gate between the two nuclear spins as %
\begin{equation}
  U_M=\left\langle M \right\vert e^{-i\frac{\pi}{4}\left[\hat{s}_{x}^{\left(\text{L}\right)}+\hat{s}_{x}^{\left(\text{R}
  \right)}\right]} e^{-i\frac{\pi}{2}\left[\hat{s}_{z}^{\left(\text{L}\right)}\hat{I}_{z}^{\left(\text{L}\right)}+\hat{s}_{z}
  ^{\left(\text{R}\right)}\hat{I}_{z}^{\left(\text{R}\right)}\right]} e^{-i\frac{\pi}{4}\hat{s}_{x}^{\left(\text{L}\right)}}\ket{\Phi^{-}},
\end{equation}
corresponding to the measurement basis $\ket{M}=\left\vert{++}\right\rangle$, $\ket{+-}$, $\ket{-+}$, $\ket{--}$.
It can be verified that the above unitary transformation is equivalent to a controlled-phase gate up to local
operations~\cite{SI}. Based on such an implementation of a controlled-phase gate between two nuclear spins,
we  propose a scalable architecture for quantum information processing including an array of diamond
nanopillars (containing NV centers) and  carbon nanotube DQDs, see Fig.~\ref{Fig4}. As an example, by controlling the positions of the diamond nanopillars, it is possible to implement controlled-phase gates as
required to generate two-dimensional (2D) cluster state efficiently~\cite{SI} in a reasonable number of steps.
We remark that local measurements on 2D cluster state are sufficient for universal measurement-based quantum computing~\cite{RN3}.
\section{Conclusion}

In conclusion, we present a hybrid quantum system consisting of NV centers and
carbon-nanotube DQDs. We show that, due to the Pauli exclusion principle, the electrons in the carbon
nanotubes are blocked in one specific triplet-like state, while NV-center electron-spin pairs evolve into a highly
entangled steady state, even under the influence of magnetic and electric noise. Considering the DQDs as the
NV center environment, this scheme can be viewed as an interesting case of quantum reservoir engineering.
By employing this steady-state entanglement between the NV-center electron spins, we propose a scalable
strategy to create cluster states in the nuclear spins, which represent a universal resource for measurement-based quantum computing. The results demonstrate that our scheme provides a promising platform for
generating entanglement between spatially separated NV centers in a deterministic way, and offers a new way towards scalable solid-state spin based quantum computing.

\section*{Acknowledgments}
This work is supported by the National Natural Science Foundation of China (11874024, 11574103, 11690032), the National Key R$\&$D Program of China (2018YFA0306600), the Fundamental Research Funds for the Central Universities. W.S. is also supported by the Postdoctoral
Innovation Talent Program, H.L. is supported by the Young Scientists Fund of the National Natural Science Foundation of China (Grant No. 11804110), and R.B. by the China Postdoctoral
Science Foundation (Grant No. 2017M622398).

\section*{Author contribution}
J.-M. Cai proposed and designed the project. W.-L. Song and T.-Y. Du carried out the calculations under the guidance of J.-M. Cai. J.-M. Cai, Ralf Betzholz, H.-B. Liu and W.-L. Song contributed to the writing of the manuscript. All authors discussed the results and commented on the manuscript.

\begin{appendix}

\section{Steady-state entanglement generation}

In our scheme for scalable solid-state spin based quantum computing, the basic building block for entanglement generation is a hybrid system consisting of two nitrogen-vacancy (NV) centers and a carbon nanotube double quantum dot (DQD). Within each NV center, the electron spin couples to the quantum dot through magnetic dipole-dipole coupling and interacts with the associated $^{15}$N nuclear spin via hyperfine interaction. In this section, we present details on the theoretical framework for the generation of steady-state entanglement between the NV-center electron spins.

\subsection{Individual subsystems}

For the electron spin of the $j$th NV center ($j=\text{L,R}$), we consider its spin-$1$ ground state with a zero-field splitting $D/2\pi=2.87\text{ GHz}$, where the degeneracy between the sublevels $\left\vert m_{\text{s}}=\pm1\right\rangle_{j}$ can be lifted by an external magnetic field $\mathbf{B}$. The system is coherently driven by adiabatic passage with optical control~\cite{Golter2014112}, which is described by the Hamiltonian
\begin{equation}\label{Hnv}
  \hat{H}_{\text{nv}}^{\left(j\right)} =\hbar D\left(\hat{S}_{z}^{\left(j\right)}\right)^{2}+g_{\text{s}}\mu_{\text{B}}\mathbf{B}\cdot\hat{\mathbf{S}}_{j} +\hbar\Omega_{j}\cos{\left(\omega_{0}t\right)}\left[\left\vert+1\right\rangle_{j}\left\langle -1\right\vert+\left\vert-1\right\rangle_{j}\left\langle+1\right\vert\right],
\end{equation}
where $\hat{\mathbf{S}}_{j}=\left(\hat{S}_{x}^{\left(j\right)},\hat{S}_{y}^{\left(j\right)},\hat{S}_{z}^{\left(j\right)}\right)$ represents the spin-1 operators and the symmetry axis of the $j$th NV center determines the $z$-axis. Furthermore, $g_{\text{s}}$ denotes the electron $g$ factor, $\mu_{\text{B}}$ is the Bohr magneton, $\Omega_{j}$ is the effective Rabi frequency of the driving field, and $\omega_{0}=2g_{s}\mu_{\text{B}}B_{z}/\hbar$ is the driving frequency.

\vspace{0.1in}

For an electron of the $j$th quantum dot ($j=\text{L,R}$), there are both spin and valley degrees of freedom contributing to the fourfold occupations in the ground shell. In an external magnetic field $\mathbf{B}$, the Hamiltonian reads~\cite{Flensberg2010,LiY2014,Szechenyi2015}
\begin{eqnarray}
  \hat{H}_{\text{qd}}^{\left( j \right)} &=& -\frac{1}{2}\Delta_{\text{SO}}\left(\mathbf{n}_{j}\cdot\hat{\boldsymbol{\sigma}}_{j} \right)\hat{\tau}_{3}^{\left( j \right)}-\frac{1}{2}\Delta_{\text{KK}'}\left( \hat{\tau}_{1}^{\left( j \right)}\cos{\varphi}+\hat{\tau}_{2}^{\left( j \right)}\sin{\varphi }\right) \nonumber\\
   & & +\frac{1}{2}g_{\text{s}}\mu_{\text{B}}\mathbf{B}\cdot\hat{\boldsymbol{\sigma}}_{j}+g_{\text{orb}}\mu_{\text{B}}\left(\mathbf{B}\cdot\mathbf{n}_{j}\right)\hat{\tau}_{3}^{\left(j\right)},
\end{eqnarray}
where $\hat{\boldsymbol{\sigma}}_{j}=\left(\hat{\sigma}_{x}^{\left(j\right)},\hat{\sigma}_{y}^{\left(j\right)},\hat{\sigma}_{z}^{\left(j\right)}\right)$ and $\hat{\boldsymbol{\tau}}_{j}=\left(\hat{\tau}_{1}^{\left(j\right)},\hat{\tau}_{2}^{\left(j\right)},\hat{\tau}_{3}^{\left(j\right)}\right)$ are the Pauli vectors of spin and valley, respectively, $\Delta_{\text{SO}}$ is the spin-orbit coupling strength~\cite{Kuemmeth2008}, $\Delta_{\text{KK}'}$ and $\varphi$ are the magnitude and phase of valley mixing~\cite{Palyi2010}, $g_{\text{s}}$ and $g_{\text{orb}}$ are the spin and orbital $g$ factors, and $\mathbf{n}_{j}=\left(\cos\alpha_{j},0,\sin\alpha_{j}\right)$ is a local tangent unit vector of the nanotube with a tilting angle $\alpha_{j}$. In the case $\alpha_{j}=0$ and $\mathbf{B}=0$, the four eigenstates form two Kramers doublets, which are separated by an energy gap $\sqrt{\Delta_{\text{SO}}^{2}+\Delta_{KK'}^{2}}$, as
\begin{eqnarray}
    & &\left\vert\Uparrow^{*}\right\rangle_{j} = -\cos{\left(\zeta/2\right)}\left\vert K'\right\rangle_{j}\left\vert\tilde{\downarrow}\right\rangle_{j}+\sin{\left(\zeta/2\right)}\left\vert K\right\rangle_{j}\left\vert\tilde{\downarrow}\right\rangle_{j},\\
    & &\left\vert\Downarrow^{*}\right\rangle_{j} = -\sin{\left(\zeta/2\right)}\left\vert K'\right\rangle_{j}\left\vert\tilde{\uparrow}\right\rangle_{j}+\cos{\left(\zeta/2\right)}\left\vert K\right\rangle_{j}\left\vert\tilde{\uparrow}\right\rangle_{j},
\end{eqnarray}
and
\begin{eqnarray}
    & &\left\vert\Uparrow\right\rangle_{j} = \cos{\left(\zeta/2\right)}\left\vert K'\right\rangle_{j}\left\vert\tilde{\uparrow}\right\rangle_{j}+\sin{\left(\zeta/2\right)}\left\vert K\right\rangle_{j}\left\vert\tilde{\uparrow}\right\rangle_{j},\\
    & &\left\vert\Downarrow\right\rangle_{j} = \sin{\left(\zeta/2\right)}\left\vert K'\right\rangle_{j}\left\vert\tilde{\downarrow}\right\rangle_{j}+\cos{\left(\zeta/2\right)}\left\vert K\right\rangle_{j}\left\vert\tilde{\downarrow}\right\rangle_{j},
\end{eqnarray}
with $\tan\zeta=\Delta_{KK'}/\Delta_{\text{SO}}$. Here, $\left\vert\tilde{\uparrow}\right\rangle_{j}$ ($\left\vert\tilde{\downarrow}\right\rangle_{j}$) and $\left\vert K'\right\rangle_{j}$ ($\left\vert K\right\rangle_{j}$) are the positive (negative) projections of $\hat{\sigma}_z$ and $\hat{\tau}_3$ respectively. For simplicity, we have set $\varphi=0$. Each Kramers doublet can serve as a valley-spin (VS) qubit \cite{Laird2013}. In our model, we focus on the lower one described by the Hamiltonian~\cite{Laird2015,Szechenyi2015,Szechenyi2017}
\begin{equation}\label{Hvs}
    \hat{H}_{\text{vs}}^{\left(j\right)}=\frac{\mu_B}{2}\mathbf{B}_{\text{eff}}^{(j)}
  \cdot\hat{\mathbf{v}}_{j},
\end{equation}
where $\hat{\mathbf{v}}_{j}=\left(\hat{v}_{x}^{\left(j\right)},\hat{v}_{y}^{\left(j\right)},\hat{v}_{z}^{\left(j\right)}\right)$ is the Pauli vector. The effective magnetic fields acting on the valley-spin qubits are given by $\mathbf{B}_{\text{eff}}^{(j)}=\mathbf{g}_{j}\cdot\mathbf{B}$, with the anisotropic $g$ tensors \begin{equation}\label{g}
  \mathbf{g}_{j}=
  \begin{pmatrix}
    g_{\parallel}\cos^{2}{\alpha_{j}}+g_{\perp}\sin^{2}{\alpha_{j}} & 0 & \left(g_{\parallel}-g_{\perp}\right)\sin{\alpha_{j}}\cos{\alpha_{j}} \\
    0 & g_{\perp} & 0 \\
    \left(g_{\parallel}-g_{\perp}\right)\sin{\alpha_{j}}\cos{\alpha_{j}} & 0 & g_{\parallel}\sin^{2}{\alpha_{j}}+g_{\perp}\cos^{2}{\alpha_{j}}
  \end{pmatrix},
\end{equation}
whose local principal values are $g_{\parallel}=g_{\text{s}}+2g_{\text{orb}}\cos{\zeta}$, $g_{\perp}=g_{\text{s}}\sin{\zeta}$.

\subsection{Interaction between subsystems}

The interaction between the $j$th NV-center electron spin and the $j$th valley-spin qubit in the carbon nanotube quantum dot can be described by the magnetic dipole-dipole interaction
\begin{equation}\label{Hee}
  \hat{H}_{\text{ee}}^{\left(j\right)}=\frac{\mu_{0}}{4\pi r^{3}_{j}}\left[\left(\mathbf{m}_{\text{vs}}^{\left(j\right)}\cdot \mathbf{m}_{\text{nv}}^{\left(j\right)}\right)-3\left(\mathbf{m}_{\text{vs}}^{\left(j\right)}\cdot\hat{\mathbf{r}}_{j}\right)\left(\mathbf{m}_{\text{nv}}^{\left(j\right)}\cdot\hat{\mathbf{r}}_{j}\right)\right],
\end{equation}
where $\mathbf{m}_{\text{vs}}^{\left(j\right)}=-\mu_{\text{B}}\mathbf{g}_{j}\cdot\hat{\mathbf{v}}_{j}/2$ and $\mathbf{m}_{\text{nv}}^{\left(j\right)}=-g_{\text{s}}\mu_{\text{B}}\hat{\mathbf{S}}_{j}$ are the magnetic moments of the valley-spin qubit and the NV-center electron spin, respectively, $\hat{\mathbf{r}}_{j}$ is the unit vector connecting them, $r_{j}$ is their distance, and $\mu_{0}$ is the magnetic constant.

\vspace{0.1in}

Due to the Coulomb blockade~\cite{Kouwenhoven2001,Hanson2007}, under a large bias voltage the electrons in the carbon nanotube transport from source to drain through the DQD via the cycle $\left(0,1\right)\rightarrow\left(1,1\right)\rightarrow\left(0,2\right)\rightarrow\left(0,1\right)$, where $\left(n_{\textbf{L}},n_{\textbf{R}}\right)$ represents the number of confined electrons in the left and right quantum dots. However, when two electrons in the $\left(1,1\right)$ configuration occupy the triplet-like states $\left\vert T_{\pm,0}\right\rangle$, the $\left(1,1\right)\rightarrow\left(0,2\right)$ transition is forbidden due to the Pauli exclusion principle~\cite{Hanson2007}. In such a Pauli-blockade regime~\cite{Szechenyi2015}, the spin-conserving tunneling between the two dots is described by
\begin{equation}
  \hat{H}_{\text{t}}=\hbar J\left(\left\vert S_{\text{g}}\right\rangle\left\langle S\right\vert +\left\vert S\right\rangle\left\langle S_{\text{g}}\right\vert\right),
\end{equation}
with the tunneling rate $J$, where $\left\vert S\right\rangle$ and $\left\vert S_{\text{g}}\right\rangle$ are the singlet-like states in the $\left(1,1\right)$ and $\left(0,2\right)$ configurations, respectively.

\subsection{Effective Hamiltonian}

The total Hamiltonian of the DQD and NV-center electron spin hybrid system can be written as
\begin{equation}
    \hat{\mathcal{H}}=\hat{H}_{\left(0,1\right)}\oplus\left(\hat{H}_{\left(1,1\right)}\oplus\hat{H}_{\left(0,2\right)}+\hat{H}_{\text{t}}\right),
\end{equation}
with
\begin{eqnarray}
    &&\hat{H}_{\left(0,1\right)} = \sum_{j=\text{L,R}}\hat{H}_{\text{nv}}^{\left(j\right)}+\hat{H}_{\text{vs}}^{\left(\text{R}\right)}+\hat{H}_{\text{ee}}^{\left(\text{R}\right)}, \\
    &&\hat{H}_{\left(1,1\right)} = \sum_{j=\text{L,R}}\left[\hat{H}_{\text{nv}}^{\left(j\right)}+\hat{H}_{\text{vs}}^{\left(j\right)}+\hat{H}_{\text{ee}}^{\left(j\right)}\right], \\
    &&\hat{H}_{\left(0,2\right)} = \sum_{j=\text{L,R}}\hat{H}_{\text{nv}}^{\left(j\right)}+\Delta\left\vert S_{\text{g}}\right\rangle\left\langle S_{\text{g}}\right\vert,
\end{eqnarray}
where $\Delta$ is the energy difference between the singlet-like states in the $\left(1,1\right)$ and $\left(0,2\right)$ configurations. We assume that the external magnetic field is $\mathbf{B}=\left(0,0,B_{z}\right)$ and the $j$th NV center is positioned in the direction $\hat{\mathbf{r}}_{j}=\left(0,0,1\right)$. After a rotating-wave approximation, the Hamiltonians $\hat{H}_{\text{nv}}^{\left(j\right)}$ and $\hat{H}_{\text{ee}}^{\left(j\right)}$ lead to the effective Hamiltonian
\begin{eqnarray}
  &&\hat{H}_{\text{es}}^{\left(j\right)} = \left(\hbar\Omega_{j}/2\right)\hat{s}_{x}^{\left(j\right)}, \\
  &&\hat{H}_{\text{dd}}^{\left(j\right)} = -\kappa_j(\tilde{\mathbf{n}}_j\cdot\hat{\mathbf{v}}_{j})\hat{s}_{z}^{(j)},
\end{eqnarray}
where $\hat{\mathbf{s}}_{j}=\left(\hat{s}_{x}^{\left(j\right)},\hat{s}_{y}^{\left(j\right)},\hat{s}_{z}^{\left(j\right)}\right)$ is the Pauli vector of the qubit encoded in the NV-center electron spin sublevels of the ground state manifold as $\left\vert0\right\rangle_{j}\equiv\left\vert m_{\text{s}}=+1\right\rangle_{j}$ and $\left\vert1\right\rangle_{j}\equiv\left\vert m_{\text{s}}=-1\right\rangle_{j}$. Here, we introduce the two vectors $\tilde{\mathbf{n}}_\text{L}=\left(-\eta,0,\xi\right)$ and $\tilde{\mathbf{n}}_\text{R}=\left(\eta,0,\xi\right)$,
with $\xi=g_{\parallel}\sin^{2}{\alpha}+g_{\perp}\cos^{2}{\alpha}$ and $\eta=\left(g_{\parallel}-g_{\perp}\right)\sin{\alpha}\cos{\alpha}$. Furthermore, we define the dipole-dipole coupling strength $\kappa_{j}=\mu_{0}\mu_{B}^{2}g_{\text{s}}/(4\pi r^{3}_{j})$. The Hamiltonian $\hat{H}_{\text{vs}}^{\left(j\right)}$ can then be rewritten as
\begin{equation}
  \hat{H}_{\text{vs}}^{\left(j\right)}=\epsilon\tilde{\mathbf{n}}_{j}\cdot\hat{\mathbf{v}}_{j},
\end{equation}
with $\epsilon=\mu_B B_z/2$.

\subsection{Unique decoupled state}

Inspired by the $\left\vert T_{0}\right\rangle$ blockade mechanism, we consider the condition $-\alpha_{\text{L}}=\alpha_{\text{R}}=\alpha$. In this case, we can explicitly write the effective Hamiltonian $\hat{H}_{\left(1,1\right)}$ (as shown Eq.~\textcolor{red}{5} in the main text) in the basis $\left\{\left\vert T_{+}\right\rangle,\left\vert T_{-}\right\rangle,\left\vert T_{0}\right\rangle,\left\vert S\right\rangle\right\}_{\mbox{DQD}}$ $\otimes$ $\left\{\left\vert\Phi^{+}\right\rangle,\left\vert\Phi^{-}\right\rangle,\left\vert\Psi^{+}\right\rangle,\left\vert\Psi^{-}\right\rangle\right\}_{\mbox{NV}}$
with the parameters $\xi'=2\xi=2\left(g_{\perp}\cos^{2}{\alpha}+g_{\parallel}\sin^{2}{\alpha}\right)$, $\eta'=\sqrt{2}\eta=\sqrt{2}\left(g_{\parallel}-g_{\perp}\right)\cos{\alpha}\sin{\alpha}$, $\kappa_{\pm}=\left(\kappa_{\text{L}}\pm\kappa_{\text{R}}\right)/2$ and $\Omega_{\pm}=\left(\Omega_{\text{L}}\pm\Omega_{\text{R}}\right)/2$. Under the conditions of $\kappa_{\text{L}} = \kappa_{\text{R}}$ and $\Omega_{\text{L}} = \Omega_{\text{R}}$, one finds that $\left\vert T_{0} \right\rangle \otimes \left\vert \Phi^{-} \right\rangle$ is the only eigenstate of $\hat{H}_{\left(1,1\right)}$ which is decoupled from the other basis states (i.e., the matrix elements on the tenth column and row are all 0 in the above Hamiltonian). During the tunnelling process, the state $\left\vert T_{0} \right\rangle \otimes \left\vert \Phi^{-} \right\rangle$ thus becomes the steady state of the total system.

\subsection{Quantum transport master equation}

In order to investigate the dynamical behavior of the entanglement generation, we use the quantum transport master equation~\cite{Gurvitz1996,LiXQ2005}
\begin{equation}
    \frac{\partial}{\partial t}\hat\rho=-\frac{i}{\hbar}\left[\hat{\mathcal{H}},\hat\rho\right]+\mathcal{L}\hat\rho,
\end{equation}
to describe the evolution of the system density operator $\hat\rho$, with the superoperator
\begin{eqnarray}
  \mathcal{L}\hat\rho=\sum_{\psi}\bigg[&&\frac{\Gamma_{\text{in}}}{2}\left(2\hat{a}_{1\psi}^{\dag}\hat\rho\hat{a}_{1\psi} -\hat{a}_{1\psi}\hat{a}_{1\psi}^{\dag}\hat\rho-\hat\rho\hat{a}_{1\psi}\hat{a}_{1\psi}^{\dag}\right)\nonumber\\
&&+\frac{\Gamma_{\text{out}}}{2}\left(2\hat{a}_{2\psi}\hat\rho\hat{a}_{2\psi}^{\dag}-\hat{a}_{2\psi}^{\dag}\hat{a}_{2\psi}\hat\rho-\hat\rho\hat{a}_{2\psi}^{\dag}\hat{a}_{2\psi}\right)\bigg],
\end{eqnarray}
where $\hat{a}_{1\psi}^{\dag}$ is the creation operator representing the injection of an unpolarized electron from the $\left(0,1\right)$ to the $\left(1,1\right)$ configuration with rate $\Gamma_{\text{in}}$ and $\hat{a}_{2\psi}$ is the annihilation operator representing the ejection of an unpolarized electron from the $\left(0,2\right)$ to the $\left(0,1\right)$ subspace with rate $\Gamma_{\text{out}}$, where $\left\{\left\vert\psi\right\rangle\right\}$ can be any complete and orthogonal set of states of the valley-spin qubits.
With the knowledge of $\hat\rho$, one can obtain the leakage current defined as~\cite{Gurvitz1996,LiXQ2005}
\begin{equation}
I=e\Gamma_{\text{out}}\sum_{\psi}\text{Tr}\big[\hat{a}_{2\psi}^{\dag}\hat{a}_{2\psi}\hat\rho\big].\end{equation}
The entanglement of two NV-center electron spins can be quantified using the concurrence, which is defined as~\cite{Hill1997,Wootters1998,Horodecki2009}
\begin{equation}
    C=\text{max}\left\{0,2\lambda_{1}-\sum_{j=1}^{4}{\lambda_{j}}\right\},
\end{equation}
where $\left\{\lambda_{1},\lambda_{2},\lambda_{3},\lambda_{4}\right\}$ are the square roots of the eigenvalues of ${  \sqrt{\tilde{\rho}}\left(\hat{s}_{\text y}\otimes \hat{s}_{\text {y}}\right)\tilde{\rho}^{\ast}\left(\hat{s}_{\text{y}}\otimes \hat{s}_{\text {y}}\right)\sqrt{\tilde{\rho}}}$ sorted in a descending order and $\tilde{\rho}$ is the reduced density operator of the NV-center electron spins by partially tracing $\hat\rho$ over the DQD degrees of freedom.

\begin{figure}
  \centering
  \includegraphics[width=8cm]{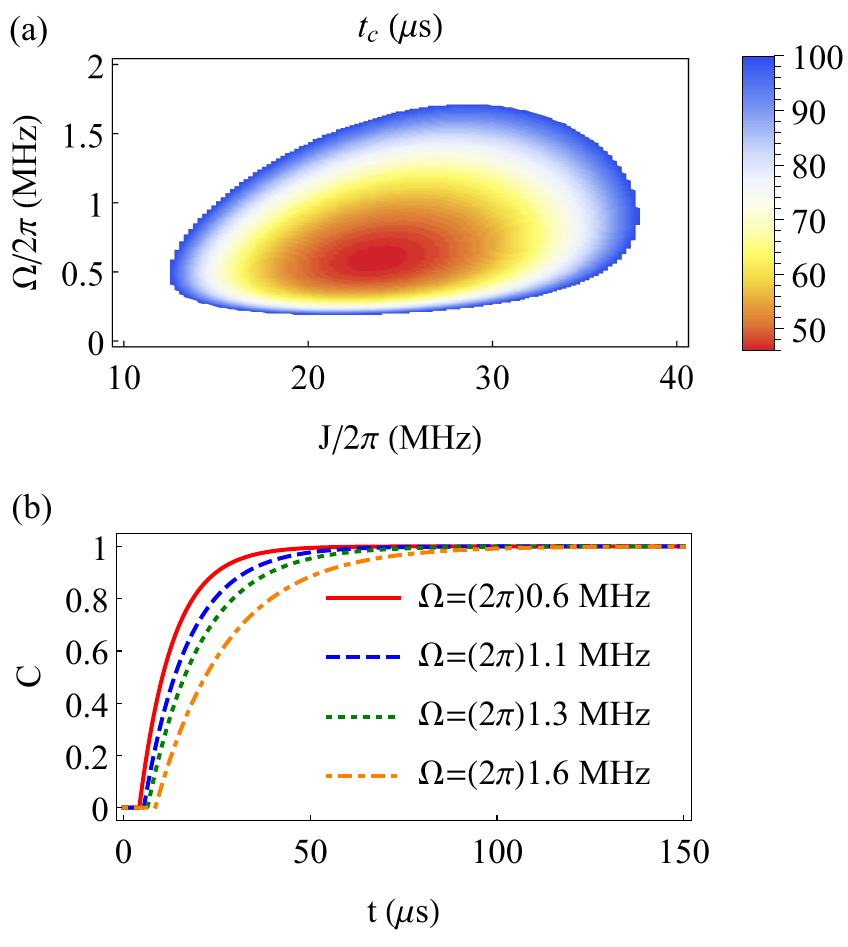}
  \caption{({\bf a}) The required time $t_{\text{c}}$ to prepare the NV-center electron spins into the maximally entangled state $\left\vert\Phi^{-}\right\rangle$ as a function of the tunneling rate $J$ and the driving Rabi frequency $\Omega$. ({\bf b}) Concurrence $C$ as a function of the evolution time $t$ for different driving Rabi frequencies $\Omega$. The remaining parameters are the ones from Fig.~\textcolor{red}{2} in the main text.}\label{Figs1}
\end{figure}

\subsection{Optimization of parameters}

The Pauli-blockade mechanism together with the magnetic dipole-dipole interaction between the NV-center electron spins and the DQD leads to the fact that the system is driven into a steady state, in which the NV-center electron spins are maximally entangled. As shown in Fig.~\textcolor{red} {1(b)} of the main text, the external magnetic field induces transitions between the states $\left\vert T_{\pm}\right\rangle$ and $\left\vert S \right\rangle$ of the DQD at rate $\sqrt{2}\epsilon\eta/\hbar$, and the microwave driving field induces transitions of the NV-center electron spins between the states $\left\vert\Phi^{+}\right\rangle$ and $\left\vert\Psi^{+}\right\rangle$ at a rate $\Omega$. On the other hand, the magnetic dipole-dipole coupling induces transitions of the hybrid system between the states $\left\vert T_{0}\right\rangle\otimes\left\vert\Psi^{\pm}\right\rangle$ and $\left\vert T_{\pm}\right\rangle\otimes\left\vert\Psi^{\mp}\right\rangle$ at rate $\sqrt{2}\kappa\eta/\hbar$. The required time $t_{\text{c}}$ to prepare the NV-center electron spins into the maximally entangled state $\left\vert\Phi^{-}\right\rangle$ depends on these parameters. In Fig.~\ref{Figs1} (a) we show that the time $t_{\text{c}}$ critically depends on the driving Rabi frequency $\Omega$ and the tunneling rate $J$. The optimized time $t_{\text{c}}=45$~$\mu$s can be achieved by choosing $J/2\pi=24$~MHz and $\Omega/2\pi=0.6$~MHz, where $\Omega\simeq \sqrt{2}\kappa\eta/\hbar$.
\section{Discussion on experimental imperfections}

In this section, we provide detailed discussions about the influence of experimental imperfections on the mechanism of steady-state entanglement generation. These experimental imperfections include magnetic field noise on the NV-center electron spins, electric potential fluctuations of the gate voltage, and the uncertainty in the depth of the NV centers. For the magnetic field noise, one can use isotopically purified diamond and dynamical decoupling techniques to reduce its influence to a large extent. In the following, we will focus on the influence of the electric potential fluctuation and the uncertainty in the positioning of the NV centers.
\subsection{Electric potential fluctuation}

\begin{figure}
  \centering
  \includegraphics[width=8cm]{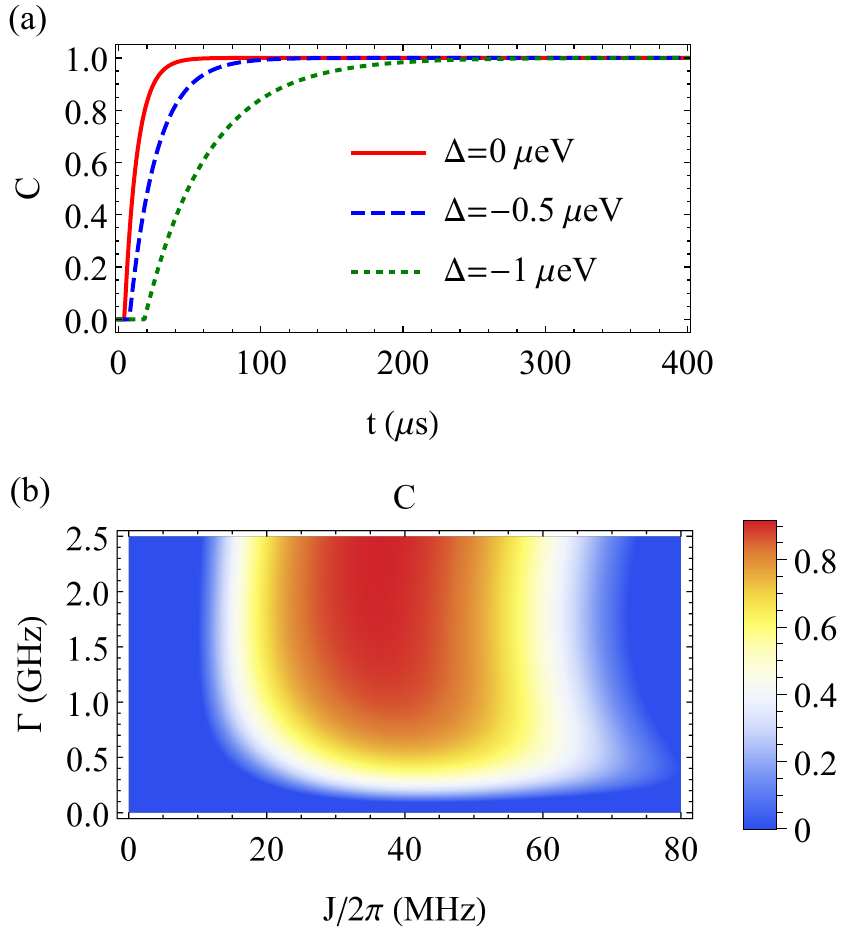}
  \caption{({\bf a}) Concurrence $C$ as a function of the evolution time $t$ for different energy shifts $\Delta$. ({\bf b}) Concurrence $C$ as a function of the tunneling rate $J$ and the electron transport rate $\Gamma = \Gamma_{\text{in}} = \Gamma_{\text{out}}$ at time $t=45$~$\mu$s for the energy shift $\Delta=-1$~$\mu$eV. The remaining parameters are the same as in Fig.~\textcolor{red}{2} of the main text.}\label{Figs2}
\end{figure}

%
\begin{figure*}
    \centering
    \includegraphics[width=11cm]{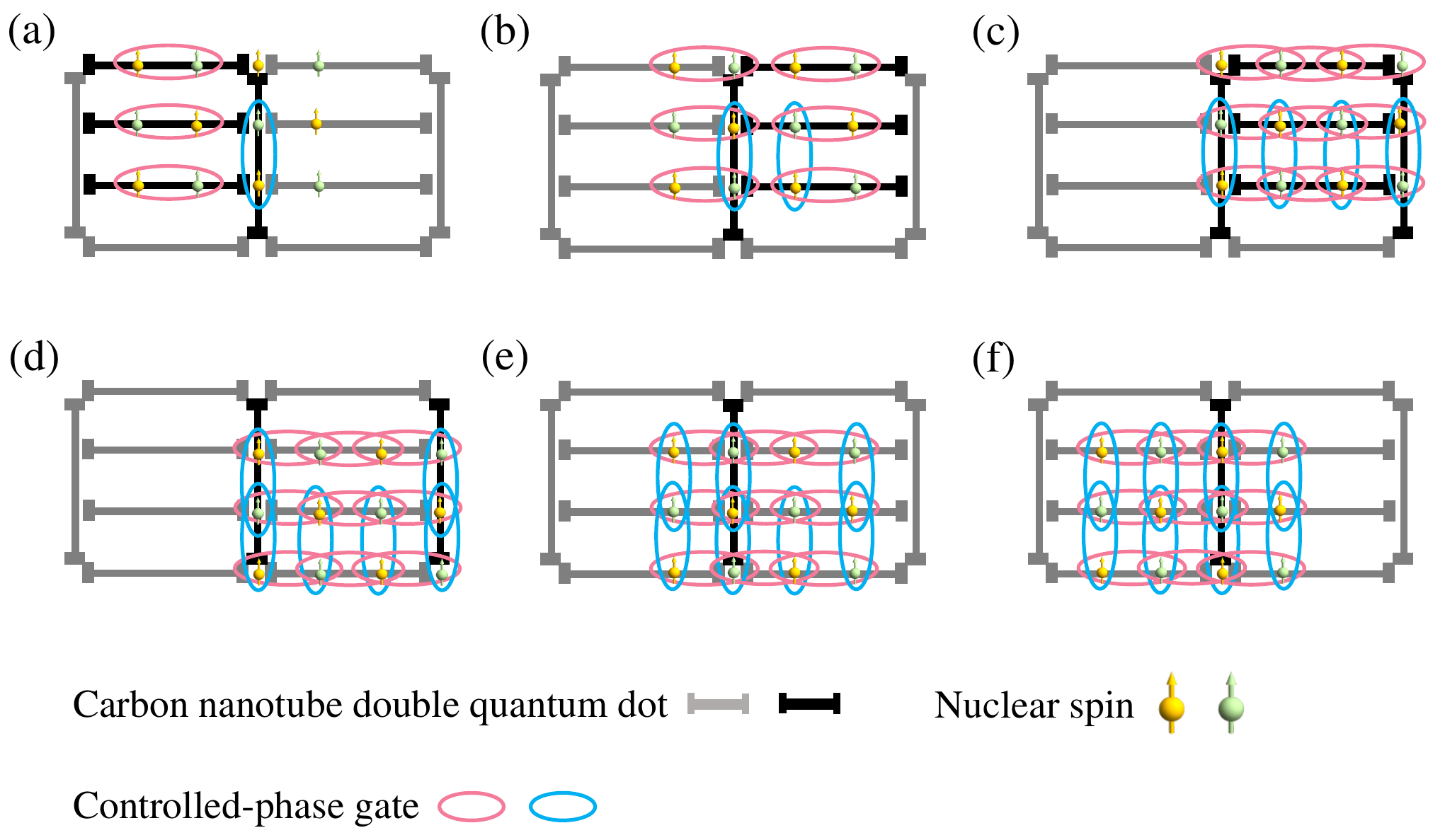}
    \caption{Schematic of the preparation of two-dimensional nuclear-spin cluster states by shifting the arrays of diamond pillars (only nuclear spins are shown for simplicity) relative to the arrays of carbon nanotubes, in the order from ({\bf a}) to ({\bf f}), and realizing controlled-phase gates mediated by the carbon nanotubes (black) sequentially on the corresponding nuclear spins in each subfigure.}
    \label{Figs3}
\end{figure*}
%
As the carbon nanotube DQD is defined by the applied gate voltage~\cite{Mason2004}, gate voltage noise will lead to electric potential fluctuations, and thereby influence the energy difference $\Delta$ between the singlet-like states $\left\vert S \right\rangle$ and $\left\vert S_{\text{g}} \right\rangle$. When the energy difference $\left\vert \Delta\right\vert > 0$, the electron tunneling between the two quantum dots will be less efficient than in the resonant case $\left\vert \Delta \right\vert = 0$, resulting in a slower generation of entanglement, as shown in Fig.~\ref{Figs2}~(a). However, this will not affect the essential mechanism for steady-state entanglement generation, namely the state $\left\vert T_{0} \right\rangle \otimes \left\vert\Phi^{-}\right\rangle$ is the only Pauli-blockade state. Thus, the maximally entangled steady-state is still achievable with a longer evolution time. Furthermore, the speed of entanglement generation can be improved by tuning the rates of the electron transport, including the injection rate $\Gamma_{\text{in}}$, the ejection rate $\Gamma_{\text{out}}$, and the tunneling rate $J$. As shown in Fig.~\ref{Figs2}~(b), the NV-center electron spins can be prepared into the maximally entangled state at the time $t=45$~$\mu$s (the same time for the ideal case without electric potential fluctuation) by choosing appropriate values of $\Gamma_{\text{in}}$, $\Gamma_{\text{out}}$, and $J$.
\subsection{Uncertainty in the positioning of NV centers}

As the magnetic dipole-dipole coupling between the $j$th NV-center electron spin and the $j$th quantum dot with strength $\kappa_{j}$ depends on the distance $r_{j}$, two NV centers doped in the diamond with different depths will lead to $\kappa_{\text{L}} \neq \kappa_{\text{R}}$. However, non-zero values of $\kappa_{-}$ would mix the state $\left\vert T_{0} \right\rangle \otimes \left\vert\Phi^{-}\right\rangle$ with other basis states and therefore degrade the entanglement in the steady state. The advanced technology of diamond scanning probes~\cite{Maletinsky_2012_NN7,Hong_2013_MB38,Schaefer-Nolte_2014_RoSI85,Rondin_2014_RoPiP77,Appel_2016_RoSI87} makes it possible to precisely control the positioning of each NV center using individual scanning probes. The problem can be further counteracted by pulsed dynamical decoupling, which has been widely used for decoherence suppression and Hamiltonian engineering. As an example, without loss of generality, we consider the case of $r_{\text{L}} \leq r_{\text{R}}$ (i.e. $\kappa_{\text{L}} \geq \kappa_{\text{R}}$).  We focus on the part of $\hat{H}_{\left(1,1\right)}$ which is related to the NV-center electron spins, namely $\hat{H}_{\text{s}}=\sum_{j=\text{L},\text{R}}\left[\left(\hbar\Omega_{j}/2\right)\hat{s}_{x}^{(j)}-\kappa_j\tilde{\mathbf{n}}_j\cdot\hat{\mathbf{v}}_{j}\hat{s}_{z}^{(j)}\right]$. By introducing an appropriate pulse sequence $\mathcal{G}=\left\{e^{i\theta\hat{s}_{y}^{\left(\text{R}\right)}/2},e^{-i\theta\hat{s}_{y}^{\left(\text{R}\right)}/2},e^{-i\theta\hat{s}_{y}^{\left(\text{R}\right)}/2},e^{i\theta\hat{s}_{y}^{\left(\text{R}\right)}/2}\right\}$ with pulse intervals $\tau_{0}$, we can engineer an effective Hamiltonian $\hat{H}'_{\text{s}}$ during the evolution time $\tau_{1}=4\tau_{0}$ which is defined by
\begin{equation}
e^{-i\tau_{1}\hat{H}'_{\mathbf{s}}/\hbar}=e^{i\theta\hat{s}_{y}^{\left(\text{R}\right)}/2}\hat{U}_{0}e^{-i\theta\hat{s}_{y}^{\left(\text{R}\right)}/2}\hat{U}_{0}e^{-i\theta\hat{s}_{y}^{\left(\text{R}\right)}/2}\hat{U}_{0}e^{i\theta\hat{s}_{y}^{\left(\text{R}\right)}/2}\hat{U}_{0}
\end{equation}
with $\hat{U}_{0}=e^{-i\tau_{0}\hat{H}_{\mathbf{s}}/\hbar}$. We note that
\begin{eqnarray}
  e^{\pm i\frac{\theta}{2}\hat{s}_{y}^{\left(\text{R}\right)}}\hat{U}_{0} e^{\mp i\frac{\theta}{2}\hat{s}_{y}^{\left(\text{R}\right)}}
  &=& \exp\left\{-\frac{i\tau_{0}}{\hbar}e^{\pm i\frac{\theta}{2}\hat{s}_{y}^{\left(\text{R}\right)}}\hat{H}_{s}e^{\mp i\frac{\theta}{2}\hat{s}_{y}^{\left(\text{R}\right)}}\right\} \nonumber\\
  &=&\exp\left\{-\frac{i\tau_{0}}{\hbar}\left[\left(\frac{\hbar\Omega_{\text{L}}}{2}\hat{s}_{x}^{(\text{L})}-\kappa_{\text{L}}\tilde{\mathbf{n}}_\text{L}\cdot\hat{\mathbf{v}}_{\text{L}}\hat{s}_{z}^{(\text{L})}\right)+\cos\theta\left(\frac{\hbar\Omega_{\text{L}}}{2}\hat{s}_{x}^{(\text{R})}-\kappa_{\text{R}}\tilde{\mathbf{n}}_\text{R}\cdot\hat{\mathbf{v}}_{\text{R}}\hat{s}_{z}^{(\text{R})}\right)\right.\right.\nonumber\\
  &&\quad\quad\quad\quad\quad\left.\left.\pm\sin\theta\left(\frac{\hbar\Omega_{\text{L}}}{2}\hat{s}_{z}^{(\text{R})}+\kappa_{\text{R}}\tilde{\mathbf{n}}_\text{R}\cdot\hat{\mathbf{v}}_{\text{R}}\hat{s}_{x}^{(\text{R})}\right)\right]\right\},
\end{eqnarray}
and thereby
\begin{eqnarray}
   e^{-\frac{i\tau_{1}\hat{H}'_{\mathbf{s}}}{\hbar}}\cong && \exp\left\{-\frac{i\tau_{1}}{\hbar}\left[\left(\frac{\hbar\Omega_{\text{L}}}{2}\hat{s}_{x}^{(\text{L})}-\kappa_{\text{L}}\tilde{\mathbf{n}}_\text{L}\cdot\hat{\mathbf{v}}_{\text{L}}^{(\text{L})}\right)\right.\right. \nonumber\\
  &&\left.\left.+\frac{1}{2}\left(1+\cos\theta\right)\left(\frac{\hbar\Omega_{\text{R}}}{2}\hat{s}_{x}^{(\text{R})}-\kappa_{\text{R}}\tilde{\mathbf{n}}_\text{R}\cdot\hat{\mathbf{v}}_{\text{R}}\hat{s}_{z}^{(\text{R})}\right)\right]\right\}.
\end{eqnarray}
By choosing proper values of $\theta$, $\Omega_{\text{L}}$ and $\Omega_{\text{R}}$ to ensure that $\Omega_{\text{L}}=\left[\left(1+\cos\theta\right)/2\right]\Omega_{\text{R}}$ and $\kappa_{\text{L}}=\left[\left(1+\cos\theta\right)/2\right]\kappa_{\text{R}}$, we are able to engineer the same effective coupling between the DQD and NV centers.%

\section{Generation of nuclear spin cluster states}

We consider $^{15}$N nuclear spins-$\frac{1}{2}$ associated with the $j$th NV center ($j=\text{L,R}$) with the Hamiltonian
\begin{equation}
    \hat{H}_{\text{ns}}^{\left(j\right)}=-g_{\text{n}}\mu_{\text{n}}\mathbf{B}\cdot\hat{\mathbf{I}}_{j},
\end{equation}
where $\hat{\mathbf{I}}_{j}=\left(\hat{I}_{x}^{\left(j\right)},\hat{I}_{y}^{\left(j\right)},\hat{I}_{z}^{\left(j\right)}\right)$ are the spin-$\frac{1}{2}$ operators, $g_{\text{n}}=0.566$ is the $g$ factor of the $^{15}$N nuclei and $\mu_{\text{n}}$ is the nuclear magneton. The hyperfine coupling between the NV-center electron spin and the $^{15}$N nuclear spin for the $j$th NV center ($j=\text{L,R}$) is given by the Hamiltonian \cite{Felton2009}
\begin{equation}
    \hat{H}_{\text{en}}^{\left(j\right)}=\hbar A_{\parallel}\hat{S}_{z}^{\left(j\right)}\hat{I}_{z}^{\left(j\right)}+\frac{\hbar A_{\perp}}{2}\left(\hat{S}_{+}^{\left(j\right)}\hat{I}_{-}^{\left(j\right)}+\hat{S}_{-}^{\left(j\right)}\hat{I}_{+}^{\left(j\right)}\right),
\end{equation}
where $\hat{S}_{\pm}^{\left(j\right)}=\hat{S}_{x}^{\left(j\right)}\pm i\hat{S}_{y}^{\left(j\right)}$, $\hat{I}_{\pm}^{\left(j\right)}=\hat{I}_{x}^{\left(j\right)}\pm i\hat{I}_{y}^{\left(j\right)}$ are the raising and lowering operators of the electron spin and the $^{15}$N nuclear spin, respectively. The coupling strength is $A_{\parallel}/{2\pi}\simeq3.03$ MHz and $A_{\perp}/{2\pi}\simeq 3.65$ MHz. Under a rotating-wave approximation, the effective hyperfine coupling Hamiltonian can be written as
\begin{equation}
  \hat{H}_{\text{hf}}=\sum_{j=\text{L,R}}\hbar A_{\parallel}\hat{s}_{z}^{\left(j\right)}\hat{I}_{z}^{\left(j\right)},
\end{equation}

In the following, we show that controlled-phase gate between$^{15}$N nuclear spins can be realized with the following four steps: (i) A $\pi/4$-$\hat{x}$ rotation on the left NV-center electron spin; (ii) Coherent evolution governed by the hyperfine interaction for time $t=\pi/2A_{\parallel}$; (iii) A $\pi/4$-$\hat{x}$ rotation on both electron spins, resulting in the following evolution operator
\begin{equation}
U_t=e^{-i\frac{\pi}{4}\left[\hat{s}_{x}^{\left(\text{L}\right)}+\hat{s}_{x}^{\left(\text{R}\right)}\right]} e^{-i\frac{\pi}{2}\left[\hat{s}_{z}^{\left(\text{L}\right)}\hat{I}_{z}^{\left(\text{L}\right)}+\hat{s}_{z}^{\left(\text{R}\right)}\hat{I}_{z}^{\left(\text{R}\right)}\right]} e^{-i\frac{\pi}{4}\hat{s}_{x}^{\left(\text{L}\right)}};
\end{equation}
(iv) Measurement of both NV-center electron spins in the $x$-basis ($\{\ket{+}$, $\ket{-}\}$) leads to an effective unitary transformation acting on the nuclear spins, as described by $U_{M}=\left\langle M \right\vert U_{t} \left\vert \Phi^{-} \right\rangle$, corresponding to the measurement basis $\ket{M}=\left\vert{++}\right\rangle$, $\ket{+-}$, $\ket{-+}$, $\ket{--}$ respectively, which can be written as
\begin{eqnarray}
&&U_{++} = -\left\vert\uparrow\uparrow\right\rangle\left\langle\uparrow\uparrow\right\vert-i\left\vert\uparrow\downarrow\right\rangle\left\langle\uparrow\downarrow\right\vert  +i\left\vert\downarrow\uparrow\right\rangle\left\langle\downarrow\uparrow\right\vert+\left\vert\downarrow\downarrow\right\rangle\left\langle\downarrow\downarrow\right\vert, \\
&&U_{+-} = -i\left\vert\uparrow\uparrow\right\rangle\left\langle\uparrow\uparrow\right\vert+\left\vert\uparrow\downarrow\right\rangle\left\langle\uparrow\downarrow\right\vert  +\left\vert\downarrow\uparrow\right\rangle\left\langle\downarrow\uparrow\right\vert-i\left\vert\downarrow\downarrow\right\rangle\left\langle\downarrow\downarrow\right\vert, \\
&&U_{-+} = i\left\vert\uparrow\uparrow\right\rangle\left\langle\uparrow\uparrow\right\vert+\left\vert\uparrow\downarrow\right\rangle\left\langle\uparrow\downarrow\right\vert  +\left\vert\downarrow\uparrow\right\rangle\left\langle\downarrow\uparrow\right\vert+i\left\vert\downarrow\downarrow\right\rangle\left\langle\downarrow\downarrow\right\vert, \\
&&U_{--} = \left\vert\uparrow\uparrow\right\rangle\left\langle\uparrow\uparrow\right\vert-i\left\vert\uparrow\downarrow\right\rangle\left\langle\uparrow\downarrow\right\vert  +i\left\vert\downarrow\uparrow\right\rangle\left\langle\downarrow\uparrow\right\vert-\left\vert\downarrow\downarrow\right\rangle\left\langle\downarrow\downarrow\right\vert,
\end{eqnarray}
in which for simplicity we use $\left\vert\uparrow\uparrow\right\rangle\equiv\left\vert\uparrow\right\rangle_{\text{L}}\left\vert\uparrow\right\rangle_{\text{R}}$, $\left\vert\uparrow\downarrow\right\rangle\equiv\left\vert\uparrow\right\rangle_{\text{L}}\left\vert\downarrow\right\rangle_{\text{R}}$, $\left\vert\downarrow\uparrow\right\rangle\equiv\left\vert\downarrow\right\rangle_{\text{L}}\left\vert\uparrow\right\rangle_{\text{R}}$, $\left\vert\downarrow\downarrow\right\rangle\equiv\left\vert\downarrow\right\rangle_{\text{L}}\left\vert\downarrow\right\rangle_{\text{R}}$.
It can be seen that the above unitary transformations are equivalent to controlled-phase gates $U_\text{CPF}$ up to local operations, namely
\begin{equation}
  U_\text{CPF}=G_{M}U_{M}=\left\vert\uparrow\uparrow\right\rangle\left\langle\uparrow\uparrow\right\vert+\left\vert\uparrow\downarrow\right\rangle\left\langle\uparrow\downarrow\right\vert+\left\vert\downarrow\uparrow\right\rangle\left\langle\downarrow\uparrow\right\vert-\left\vert\downarrow\downarrow\right\rangle\left\langle\downarrow\downarrow\right\vert,
\end{equation}
with
\begin{eqnarray}
      &&G_{++} = \left(\left\vert\uparrow\right\rangle_{\text{L}}\left\langle\uparrow\right\vert+i\left\vert\downarrow\right\rangle_\text{L}\left\langle\downarrow\right\vert\right)  \otimes\left(-\left\vert\uparrow\right\rangle_\text{R}\left\langle\uparrow\right\vert+i\left\vert\downarrow\right\rangle_\text{R}\left\langle\downarrow\right\vert\right),\\
      &&G_{+-} = \left(\left\vert\uparrow\right\rangle_{\text{L}}\left\langle\uparrow\right\vert-i\left\vert\downarrow\right\rangle_\text{L}\left\langle\downarrow\right\vert\right)  \otimes\left(i\left\vert\uparrow\right\rangle_\text{R}\left\langle\uparrow\right\vert+\left\vert\downarrow\right\rangle_\text{R}\left\langle\downarrow\right\vert\right),\\
      &&G_{-+} = \left(\left\vert\uparrow\right\rangle_{\text{L}}\left\langle\uparrow\right\vert+i\left\vert\downarrow\right\rangle_\text{L}\left\langle\downarrow\right\vert\right)  \otimes\left(-i\left\vert\uparrow\right\rangle_\text{R}\left\langle\uparrow\right\vert+\left\vert\downarrow\right\rangle_\text{R}\left\langle\downarrow\right\vert\right),\\
      &&G_{--} = \left(\left\vert\uparrow\right\rangle_{\text{L}}\left\langle\uparrow\right\vert-i\left\vert\downarrow\right\rangle_\text{L}\left\langle\downarrow\right\vert\right)  \otimes\left(\left\vert\uparrow\right\rangle_\text{R}\left\langle\uparrow\right\vert+i\left\vert\downarrow\right\rangle_\text{R}\left\langle\downarrow\right\vert\right).
\end{eqnarray}
Based on such an implementation of controlled-phase gates between nuclear spins, an array of carbon nanotubes as presented in Fig.~\textcolor{red}{4}(a-b) of the main text allows one to prepare one-dimensional nuclear-spin cluster states. Similarly, two-dimensional nuclear-spin cluster states can be generated using a lattice of carbon nanotubes in six steps, see Fig.~\ref{Figs3} for an example of a $3\times4$ cluster state generation.

\end{appendix}

\bibliography{article}

\end{document}